\documentclass[aps,floatfix,prx,superscriptaddress,reprint,showpacs,10pt]{revtex4-1}

\usepackage{lipsum}
\usepackage{color}
\usepackage{hyperref}
\usepackage{amsfonts}
\usepackage{amsmath}
\usepackage{mathtools}
\usepackage{amssymb}
\usepackage{amsthm}
\usepackage[linesnumbered, ruled]{algorithm2e}
\usepackage[caption=false]{subfig}

\newcommand*{\trace}{\mathsf{Tr}} 
\DeclareMathOperator{\Span}{span}

\newcommand{\ket}[1]{|#1\rangle}
\newcommand{\bra}[1]{\langle#1|}
\newcommand{\braket}[1]{\langle #1\rangle}

\begin{document}

\title{Efficient approximation for global functions of matrix product operators}

\begin{abstract}
Building on a previously introduced block Lanczos method, 
we demonstrate how to approximate any operator function of the form $\trace f(A)$ when the 
argument $A$ is given as a Hermitian matrix product operator.
This gives access to quantities that, depending on the full spectrum, are difficult to access for standard tensor network techniques,
such as the von Neumann entropy and the trace norm of an MPO.
We present a modified, more efficient strategy for computing thermal properties of short- or long-range Hamiltonians,
and illustrate the performance of the method
with numerical results for the thermal equilibrium states of the Lipkin-Meshkov-Glick and Ising Hamiltonians.
\end{abstract}

\author{Moritz August}
\affiliation{Department of Informatics, Technical University of Munich, 85748 Garching, Germany}
\email{august@in.tum.de}
\author{Mari Carmen Ba\~{n}uls}
\affiliation{Max-Planck Institute for Quantum Optics, 85748 Garching, Germany}
\email{banulsm@mpq.mpg.de}

\maketitle

\section{Introduction}
\label{sec:introduction}



Tensor networks (TN) have proved to be adequate ans\"atze for the description of 
ground states, low energy excitations and thermal equilibrium states of
quantum many body systems \cite{schollwock2011density,verstraete2008matrix,Orus2014a}.
Within some limitations, they can also be used to simulate real time dynamics \cite{garcia2006time,barthel2009spectral,muellerhermes2012njp}.
In particular, the matrix product state (MPS) ansatz constitutes the best understood and most used 
TN family, and it underlies the success
of the celebrated density matrix renormalization group algorithm (DMRG) \cite{White1992}.

The matrix product operator (MPO) 
generalizes MPS to operators, and
provides a variational ansatz for mixed states \cite{verstraete2004matrix,zwolak2004mixed,pirvu2010matrix}.
It can also be used to efficiently describe many Hamiltonians of physical interest,
as well as approximate evolution operators, among others,
 and has become a most
 useful tool in the application and understanding of TN algorithms in one and two dimensions.
An operator given in MPO form can be efficiently applied to a MPS 
  using standard TN techniques, and this operation 
  constitutes a building block 
  for algorithms that search for the ground state or simulate time evolution of a MPS.

The MPO form allows also an efficient calculation of some operator properties (e.g. the trace of the operator or of a few of its integer powers, which gives access to some integer $\alpha$-Renyi entropies with $\alpha\geq2$ \cite{renyi1961measures}).
However, already for moderate system sizes, it is in general not possible to access the full spectrum of the MPO,
and hence there are physical properties that are difficult to estimate. 
This includes
for instance the von Neumann entropy of a mixed state, or the trace distance between two MPOs.
These quantities have in common that they can be written as functions of Hermitian matrices $A \in \mathbb{C}^{N \times N}$, such as Hamiltonians or density operators, that for finite dimensional systems take the form $\trace f(A)$, or more generally sums and products of such terms.

We have recently introduced a numerical method~\cite{august2017approximation} that based on a particular version of the Lanczos algorithm reformulated for MPOs approximates such functions for arbitrary Hermitian inputs. 
The algorithm implicitly performs a Gauss quadrature approximation and we have shown that it converges to the exact value in the absence of approximation errors.

In this paper we demonstrate that the algorithm can be used to compute physical quantities that are difficult to access in 
standard MPO calculations. 
The method reveals to be particularly useful in the case of thermal equilibrium states, since many thermal 
properties can be written as functions of the Hamiltonian. If the latter has a MPO expression, this can be used, for instance, 
to detect thermal phase transitions.

The rest of this work is structured as follows: in section~\ref{sec:the_approximation_method} we briefly introduce our approximation method. 
Section~\ref{sec:approximation_of_gibbs_states} explicitly shows how to use the method to approximate 
observables in thermal equilibrium, as well as distances between Gibbs states.
This strategy is illustrated with numerical results in section~\ref{sec:numerical_results}.
Finally we summarize our conclusions in section~\ref{sec:conclusion}.

\section{The approximation method}
\label{sec:the_approximation_method}


Krylov type methods have already been used with success in combination with matrix product states (MPS) for the approximation of extremal eigenstates~\cite{huckle2012subspace} or time evolution~\cite{schmitteckert04,garcia2006time, wall2012out, hubig2017symmetry, keilmann2008dynamical} and dynamical correlation functions \cite{dargel12dyn} as well as spectral functions~\cite{dargel2011adaptive}. 
In general these methods, that rely on a solid mathematical theory~\cite{krylov1931numerical, lanczos1950iteration}, construct a
basis of the Krylov subspace for an input matrix $A$ and an initial vector $b$, $\mathcal{K}(A,b):=\Span \{A^0b, A^1b, A^2b,\cdots,A^{K-1}b\}$,
and a projection of $A$ onto the subspace, $T_K$.
Their implementation only requires basic linear algebra operations, such as scalar multiplication, addition and inner products of vectors, 
which allows an easy reformulation in the tensor network framework.
In particular the Lanczos algorithm,
one of the best known Krylov subspace methods, 
finds the most significant eigenvalues and eigenvectors of a Hermitian matrix.

The basic vector-based Lanczos algorithm can be generalized to matrices by considering them to be block vectors comprised of several individual column vectors, each of them corresponding to an individual Krylov subspace.
These block Lanczos methods can then construct a basis of block vectors starting from 
a usually not square initial block that plays the role of the starting vector $b$ used in the standard Lanczos algorithm, see e.g.~\cite{golub1977block, golub1981block}.
The benefit of these algorithms thereby lies in their ability to approximate several extremal eigenvalues or the solution of linear systems with multiple solution vectors simultaneously from a possibly larger search space while often yielding implementations with a better runtime compared to the repeated application of the original Lanczos algorithm. 

While most block Lanczos algorithms are based on inner products and norms for the individual column vectors constituting the block vectors, a variant, called global block Lanczos algorithms, has been developed~\cite{jbilou1999global, elbouyahyaoui2009algebraic, bellalij2015bounding} which is based on inner products and corresponding norms defined for matrices.
Based on these works, in \cite{august2017approximation} we recently presented a global block Lanczos method, from now on also simply referred to as block Lanczos method, for MPO operators
that makes use of TN techniques.
In particular we use the Hilbert-Schmidt inner product $\langle U, V \rangle = \trace \left (U^{\dagger} V\right)$ and 
the induced Frobenius norm, and choose the identity matrix as the starting point.
The algorithm then constructs a
basis for the (operator) subspace $\mathcal{K}(A)=\Span \{A^0, A^1, A^2,\cdots,A^{K-1}\}$.

Lanczos methods can be used to approximate functions of the form $b^{\dagger}f(A)b$
when numerical diagonalization is infeasible~\cite{golub1994matrices},
a result which relies on a rigorous connection to Gauss quadrature.
In the case of the global block method, this connection can be exploited to approximate functions of the form $\trace \left ( f(A)\right)$.
More precisely, using any unitary complex 
initial matrix $U$, the trace  can be written as a Riemann-Stieltjes integral,
$\trace \left [ f(A)\right]=\trace \left [U^{\dagger} f(A) U\right]=\int f(\lambda) d\mu(\lambda)$,
where the integral is over the spectrum of $A$, and the measure $\mu(\lambda)$ is a piecewise 
constant distribution depending on the choice $U$, see e.g. \cite{august2017approximation} for details.
It can be shown~\cite{bellalij2015bounding}
that the eigenvalues of the projection $T_K$ of $A$ onto
the Krylov subspace, produced by the global block Lanczos method for the initial matrix $U$, 
correspond precisely to the nodes of the $K$-node Gaussian quadrature approximation of the integral above.
The corresponding weights are contained in the respective eigenvectors.
When the function $f$ is applied to the projected matrix, the first element of the diagonal of $f(T_K)$, scaled by the squared norm of the starting matrix $b,$
evaluates the Gauss quadrature and thus provides an approximation to the trace.
However, using a fully unitary starting matrix has so far been prohibitive in practice since in this case the complexity of the algorithm is no better than that of exact diagonalization.


\begin{algorithm}[t]
\setlength{\leftskip}{10pt}
\setlength{\skiprule}{10pt}
\caption{MPO Function Approximation}
\label{approx}
    \SetKwInOut{Input}{Input}
    \SetKwInOut{Output}{Output}
    \SetKwFunction{multOpt}{multiply}
    \SetKwFunction{addOpt}{sum}
    \SetKwFunction{contract}{innerProduct}
	\SetKwFunction{scalMult}{multiply}
	\SetKwFunction{spec}{spectralDecomposition}
	\SetKwFunction{checks}{checkStop}

    \Input{MPO $A[D_{A}] \in \mathbb{C}^{N \times N}$, Starting unitary MPO $U[D_{init}] \in \mathbb{C}^{N \times N}$, Number of Dimensions $K$, Maximal Bond-Dimension $D_{max}$, Stopping Criteria $\mathcal{S}$}    
    $U_0 \leftarrow 0$ \;
    $V_0 \leftarrow U$ \;
    \For{$i \leftarrow 1 ; i \leq K $}{
	$\beta_i \leftarrow \sqrt{ \langle V_{i-1}| V_{i-1} \rangle}$ \;
    \If{$\beta_i = 0$}{break \;}
    $U_i \leftarrow  \scalMult(1 / \beta_i,V_{i-1})$ \;    
$V_i \leftarrow \multOpt(A, U_{i}, D_{max})$ \;
$V_i \leftarrow \addOpt(V_i, - \beta_i U_{i-1}, D_{max})$ \;
    $\alpha_i \leftarrow \langle U_i| V_i\rangle$ \;
$V_i \leftarrow \addOpt(V_i, -\alpha_i U_{i}, D_{max})$ \;
    
    $V_T \Lambda_T V_T^* \leftarrow \spec(T_i)$ \;
    $\mathcal{G}f \leftarrow \beta_1^2 e_1^T V_T f(\Lambda_T) V_T^* e_1$ \;
    \If{$\checks(\mathcal{G}f, \Lambda_T, \mathcal{S})$}{break \;}
    }
    
  	\Output{Approximation $\mathcal{G}f$ of $\trace f(A)$}
\end{algorithm}

The algorithm proposed in \cite{august2017approximation} and further analyzed in~\cite{august2017towards}
applies the above property to approximate global functions of an operator $A$ given in MPO form.
We show the method schematically as pseudocode in Algorithm~\ref{approx}.
As explained above, it proceeds by applying a global block Lanczos method with the identity as initial matrix.
The identity is chosen here simply because it yields an exact MPO representation with bond dimension $D=1$. 
In principle however, any unitary MPO could be used, a fact we will take advantage of in Section~\ref{sec:approximation_of_gibbs_states}. 
It is important to note here that only due to the use of the TN framework we are able to use full unitary starting matrices.
The method then constructs the orthogonal basis of the Krylov subspace in the usual way, i.e. successively applying the 
operator $A$ and orthogonalizing, but restricting all the basis matrices to be of the MPO form
with a maximum bond dimension of $D_{max}$.
Iteratively, the method thus constructs the projection matrix 
\begin{equation*}
	T_K = \left[ \begin{matrix}
		\alpha_1 & \beta_2 & & \mathbf{0}\\
		\beta_2 & \alpha_2 & \ddots & \\
		& \ddots & \ddots & \beta_K \\
		\mathbf{0} & & \beta_K & \alpha_K \end{matrix} \right]
	\end{equation*}
where the $\alpha_i$ and $\beta_i$ are computed as shown in Algorithm~\ref{approx}.
The spectrum of $T_K$ then in essence constitutes an approximation to the spectrum of the input MPO $A$.

The multiplication and the sum of MPOs, denoted as the subfunctions \texttt{sum} and \texttt{multiply} in the pseudocode, involve an optimization over MPOs
for a given target bond dimension.
In our algorithm we use the bond dimension required for an exact representation until it exceeds the maximal bond dimension $D_{max}$.
All the operations involved can be carried out efficiently for MPO operators, so that the method allows the approximation
of functions of the $A$ which would be otherwise inaccessible,
as in principle they would require to compute the whole spectrum.

Representing the basis elements as MPS, or vectorized MPOs, with limited bond dimension up to $D_{\max}$
introduces a truncation error. 
Together with the error introduced by the limited dimension of the Krylov subspace, the truncation error determines the accuracy of the 
approximation. 
Since in absence of these errors the method is known to converge to the exact result, by adjusting these parameters, the numerical errors can be controlled.

Further details about the theoretical background and the algorithm itself can be found in ~\cite{golub1994matrices, bellalij2015bounding, august2017approximation}.
In the next sections we show how this method can be applied to compute physical quantities
in quantum many-body systems.

\section{Functions of Gibbs states}
\label{sec:approximation_of_gibbs_states}

For a quantum system governed by a Hamiltonian $H$, the operator
\begin{align}
	\rho(H, \beta) = \frac{e^{-\beta H}}{Z}
\end{align}
describes the thermal equilibrium state at inverse temperature $\beta$ 
and is called Gibbs ensemble.
The denominator in the expression is the  partition function, $ Z = \trace e^{-\beta H}$.

It has been shown that MPOs can approximate Gibbs states of local Hamiltonians accurately \cite{Hastings2006a,molnar15gibbs}.
From a numerical perspective, 
MPO approximations to thermal states can be found with efficient imaginary time evolution 
algorithms~\cite{verstraete2004matrix, zwolak2004mixed, garcia2006time}.


In principle, we could then apply the block Lanczos method to the MPO approximation found with one such approximation algorithm,
in order to evaluate a given function $f(\rho)$.
However, the particular form of the Gibbs state allows us to write quantities directly as a function of the 
Hamiltonian, so that instead of approximating $f(\rho)$ as a function of $\rho$, we can approximate $f\left[\rho(H,\beta)\right]$ as a function of $H$.

This strategy offers multiple advantages.
Firstly, for many interesting physical problems, for instance for local interactions,
$H$ has an exact MPO representation.
In that case
the method does not suffer from initial approximation errors, in contrast to the case where an approximation of 
$\rho(H, \beta)$ is used as input.
In the case of a long-range Hamiltonian, often a good MPO approximation of $H$ exists with reduced bond dimension~\cite{pirvu2010matrix}. In contrast, approximating the Gibbs state as an MPO may be prohibitive
in terms of bond dimension, such that this strategy allows access to thermal observables which would be otherwise 
difficult or infeasible to approximate numerically.
A second advantage is that the (exact or approximate) MPO representation of $H$ usually has a much lower bond dimension than what is required for a good approximation of $\rho(H, \beta)$, significantly reducing the overall execution time of the algorithm.
Thirdly, since the Krylov subspace constructed by our algorithm only depends on $H$, we can in principle use it to approximate multiple different functions of $H$ 
and a broad range of different temperatures 
in a single run.
As we illustrate in the next paragraphs, this strategy can be applied to compute different physical quantities
which are not easy to estimate for a general MPO,
including the von Neumann entropy, or the trace norm distance between two
thermal states at different temperatures.
The latter quantity, as the 
equally 
accessible thermal fidelity, can be used to detect thermal phase transitions~\cite{zanardi2007mixed,quan2009quantum}.
Other quantities that can be approximated by this method include the heat capacity \cite{wilms2012finite}
and correlation functions.


\subsubsection{Thermodynamic quantities}
	The von Neumann entropy is defined as $S(\rho) = - \trace \left( \rho \ln \rho \right)$. 
		For a thermal state it can be written as a function of $H$, 
		\begin{align}
			S\left[\rho(H, \beta)\right] &= \beta \frac{F}{Z} + \ln Z.
		\end{align}
		where $F=\trace \left( e^{-\beta H}H \right)$ is the free energy.
		Thus $S$ can be expressed in terms of  two trace functions, $F$ and $Z$, that depend on the same input operator $H$.
		Therefore, as discussed above,
		we can approximate and then combine $F$ and $Z$ to yield $S$ in a single run of the algorithm.
		On the other hand, $\beta$ is just a parameter of the function, 
		so that 
		we can also approximate $S$ for arbitrarily many values of $\beta$ simultaneously in one run of the method.
		In the following, we will consider the von Neumann entropy per particle denoted as $s = S/L$ where $L$ is the system size.
	 
	We can also estimate other thermodynamic quantities that derive from the partition function, i.\ e.\ any quantity that is expressible in terms of derivatives of $Z$ or $\ln Z$.
	Next to the von Neumann entropy, one other such quantity, which is especially relevant in the context of probing thermal phase transitions, is the specific heat capacity (i.e. heat capacity per site)
		\begin{align}
			c = \frac{\beta^2}{L}\frac{\partial^2 \ln Z}{\partial \beta^2}
		\end{align}
		which can be written as
		\begin{align}
			c &= \frac{\beta^2}{L} \left[ \frac{G}{Z}  - \left(\frac{F}{Z} \right)^2 \right]
		\end{align}
		where $G=\trace \left(H^2 e^{-\beta H}\right)$. 
		We can obtain $c$ by simultaneously approximating $F$, $G$ and $Z$, 
	as in the case of 
		the von Neumann entropy.

\subsubsection{Distance measures}
The trace norm of an  operator $A$, defined
$\| A\|_1=\trace \sqrt{A^{\dagger}A}$, 
is difficult to compute for MPOs, as in general no efficient MPO description exists for the square root.
However, it can be easily
computed using the block Lanczos method, as we discussed in \cite{august2017approximation}.
If the operator is a function of the Hamiltonian, the strategy presented here will in general improve the approximation as compared to the case where an MPO representation of $A^{\dagger}A$ is used as input.

This strategy gives easy access to 
the trace distance between thermal states at different temperatures.
More concretely, for two inverse temperatures $\beta_0$ and $\beta_1$, the trace distance
\begin{align}			
D_T(\beta_0,\beta_1)= \left \| \frac{e^{-\beta_0 H}}{Z(\beta_0)}-\frac{e^{-\beta_1 H}}{Z(\beta_1)} \right \|_1
\label{eq:trdist}
\end{align}
is a function of the Hamiltonian $H$.
To approximate it with our method we need an estimate for the partition functions 
at both temperatures
$Z(\beta_0)$ and $Z(\beta_1)$.
As before, the latter can be approximated independently, and their values can then be used to evaluate the trace distance. 
Notice that the distance between two arbitrary MPO operators can be approximated by straightforward application of the method in ~\cite{august2017approximation} to the MPO that represents their difference.
Using this procedure we can compute also trace distances between thermal equilibrium states 
at the same temperature, but varying Hamiltonian parameters, whenever a MPO approximation can be found to the Gibbs states.
This however may be difficult, for instance if the Hamiltonian has long-range interactions.
	
Another distance measure 
between quantum states is the Uhlmann fidelity, 
which for a pair of states  $\rho$ and $\sigma$ is defined as
$F(\rho, \sigma)=\trace \sqrt{\rho^{1/2}\sigma \rho^{1/2}}$.
	As was shown in~\cite{zanardi2007mixed} and further studied in \cite{quan2009quantum}, 
	the \emph{thermal fidelity}, i.e. the fidelity between two thermal equilibrium states for the same Hamiltonian at different 
	temperatures, can detect a thermal phase transition.
	For two inverse temperatures $\beta_0$ and $\beta_1$, the thermal fidelity 
	can be expressed  in terms of three partition functions,
			\begin{align}
			F_T(\beta_0, \beta_1) &= \frac{Z\left( \frac{\beta_0 + \beta_1}{2} \right)}{\sqrt{Z(\beta_0)Z(\beta_1)}}, 
		\end{align}
which implies that it can be approximated with the strategy outlined in the previous section.
\subsubsection{Other expectation values}
Interestingly, the same method can also be used to approximate the expectation values of some observables.
 Generically, expectation values take the form $\trace \left [ f(A)O \right ]$ which in principle cannot be directly approximated.
However, if $O$ is positive, we can write
\begin{align}
	\trace \left [ f(A)O \right ]=\trace\left [ \sqrt{O}f(A)\sqrt{O}^{\dagger} \right ]=\int f(\lambda) d \mu_O(\lambda),
\end{align}
with a function $\mu_O$ specific for the operator $O$.
The positivity  of $O$ 
 is required to ensure that $\mu_O$ is a distribution function.
If $\sqrt{O}$ admits an efficient representation as MPO, we can compute the expression above by simply using
$\sqrt{O}^{\dagger}$ as the starting point of our algorithm, instead of the identity.
Even if the operator $O$ is not positive, it can always be written as
linear combination of positive terms, e.g. splitting it in positive and negative part, although
other decompositions with more terms may be more convenient.
If each of these positive terms has an efficient MPO representation of its square root, it will be possible to approximate the expectation value by computing 
each contribution independently.
Operators that allow this treatment, are, for instance, few-body spin correlators, 
as we detail in the next section.


\section{Numerical results}
\label{sec:numerical_results}

To illustrate the performance of the method we present numerical results for the
exploration of a thermal phase transition in a long-range model, and the extraction of two-point 
spin correlation functions.


\subsection{Phase Transition of the LMG model}


The  Lipkin-Meshkov-Glick (LMG) Hamiltonian~\cite{lipkin1965validity, meshkov1965validity, glick1965validity} for a system of $L$ spin-$1/2$ particles is given by
\begin{align}
H = -\frac{S_x^2}{L}-2hS_z,
\label{eq:LMG}
\end{align}
where $S_{\alpha} = \sum_i^L \sigma_i^{\alpha}/2$ for $\alpha \in \{x,y,z\}$
 are the collective spin operators,  and each $\sigma_i^{\alpha}$ stands for
 the corresponding Pauli operator at site $i$. 
The model
exhibits a quantum phase transition for $h=1$~\cite{botet82lmg}
and a thermal phase transition for $h<1$
at a critical temperature
\begin{align}
	T_c(h) = \frac{h}{2 \tanh^{-1}(h) }.
	\label{eq:Tc_LMG}
\end{align}
The model has been largely studied in the literature, at both zero and finite temperature.
Here we focus on the properties of the thermal equilibrium states, so
specially related to our study are some recent works that explored the thermal phase transition from a quantum information 
point of view.
The first study of the finite temperature phase diagram of the model using the thermal fidelity was performed in
\cite{quan2009quantum}.
In Wilms et.\ al.~\cite{wilms2012finite}, using both analytical and numerical results, it was demonstrated that the mutual information,
which measures quantum and classical correlations, was sensitive to the phase transition.
The fidelity metric was also used by Scherer et.\ al.~\cite{scherer2009finite} to explore analytically 
the phase transition of the isotropic LMG model.
\begin{figure}[t]
%
%
	\subfloat[$h=0.2$] {\includegraphics[width=0.25\textwidth]{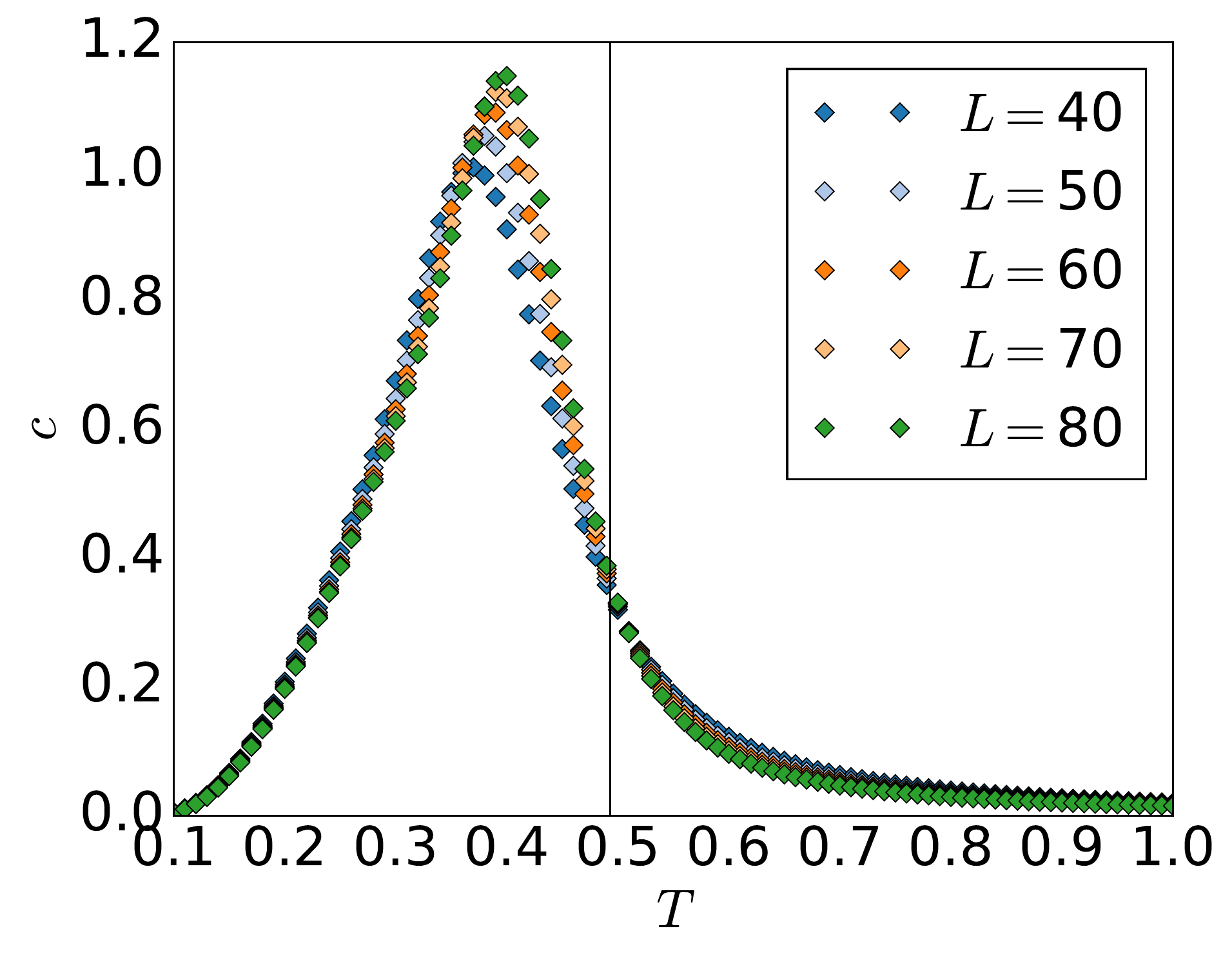}}
	\subfloat[$h=1.2$] {\includegraphics[width=0.25\textwidth]{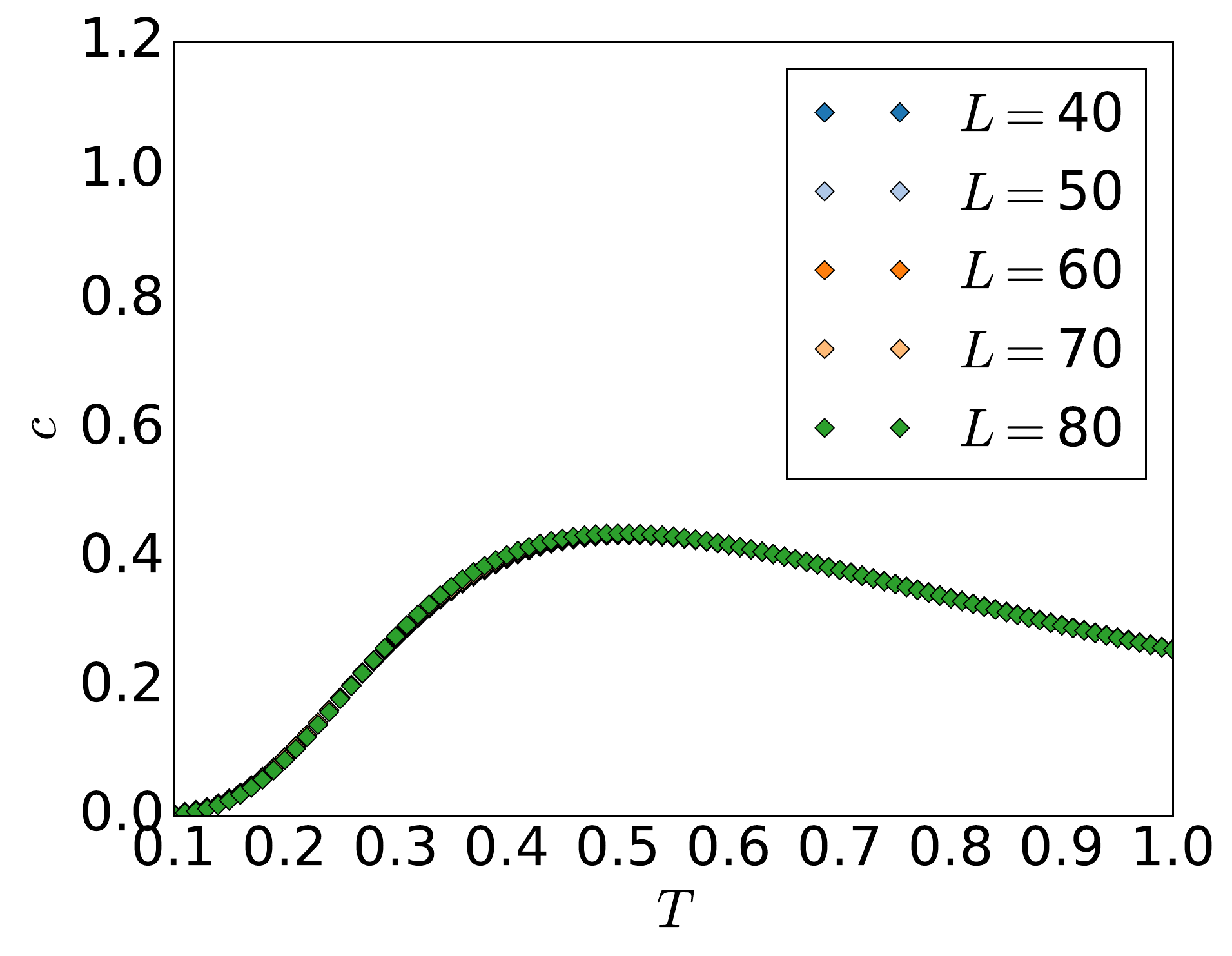}}\\
	\subfloat[Extrapolation of $T_c$] {\includegraphics[width=0.25\textwidth]{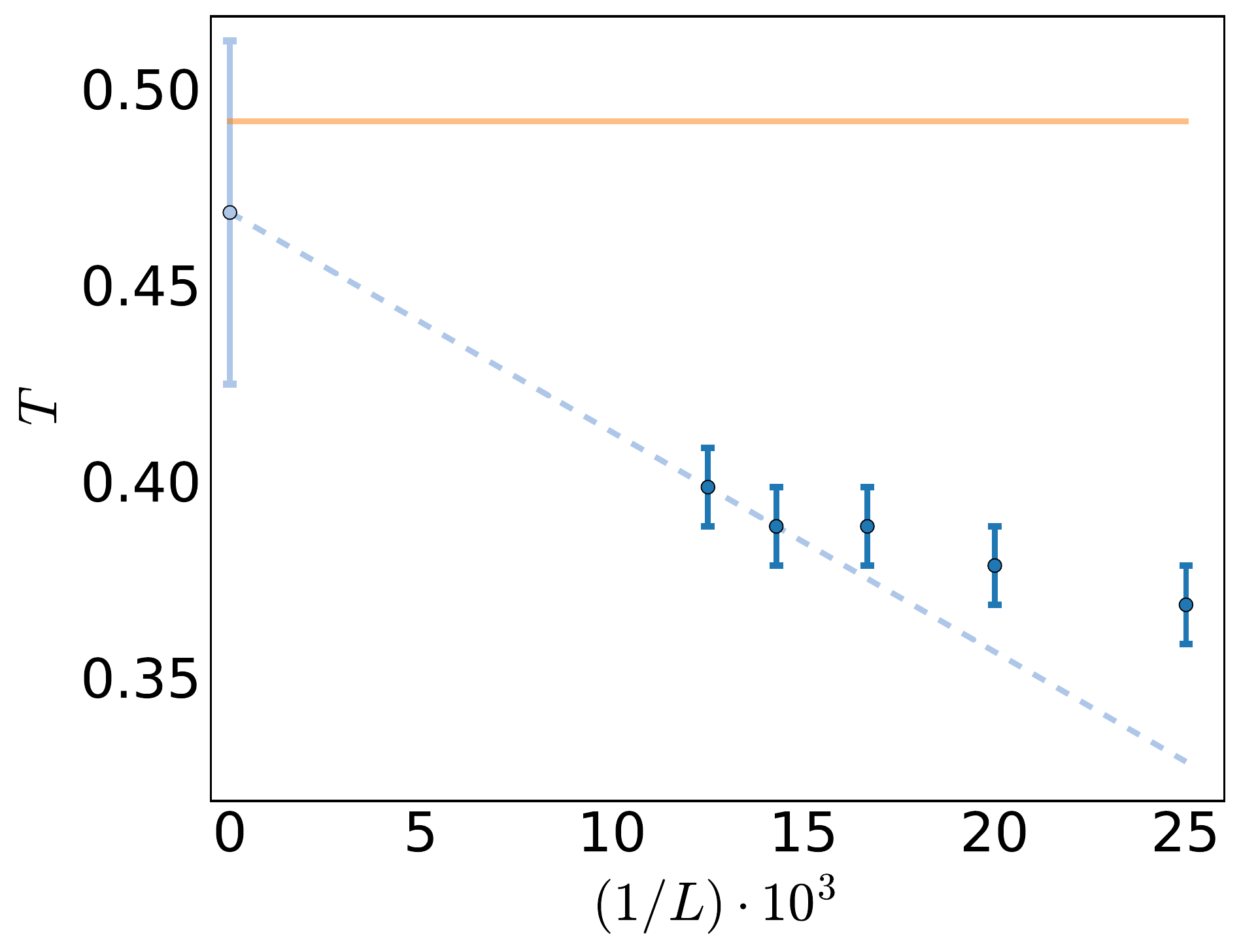}}
	\caption{
Heat capacity per particle in the thermal equilibrium state of the LMG model, for different system sizes $40\leq L\leq 80$ and 
$h=0.2$ (a) and $h=1.2$ (b) as a function of temperature. The vertical scale is the same in both plots for the sake of comparison.
		Our results are in good agreement with those presented in ~\cite{wilms2012finite}.
The black vertical line in Figure (a) indicates the critical temperature $T_c$. Figure (c) shows 
the scaling of the peak position with the system size. The error bars of the data points reflect the imprecision in the location of each peak. 
Extrapolating the results of the largest systems ($L=70,\,80$) we
estimate a critical temperature $T_c\approx0.47$.
	}
	\label{heat_cap}
\end{figure}

\begin{figure}[t]
	\center
	\subfloat[$h=0.2$] {\includegraphics[width=0.25\textwidth]{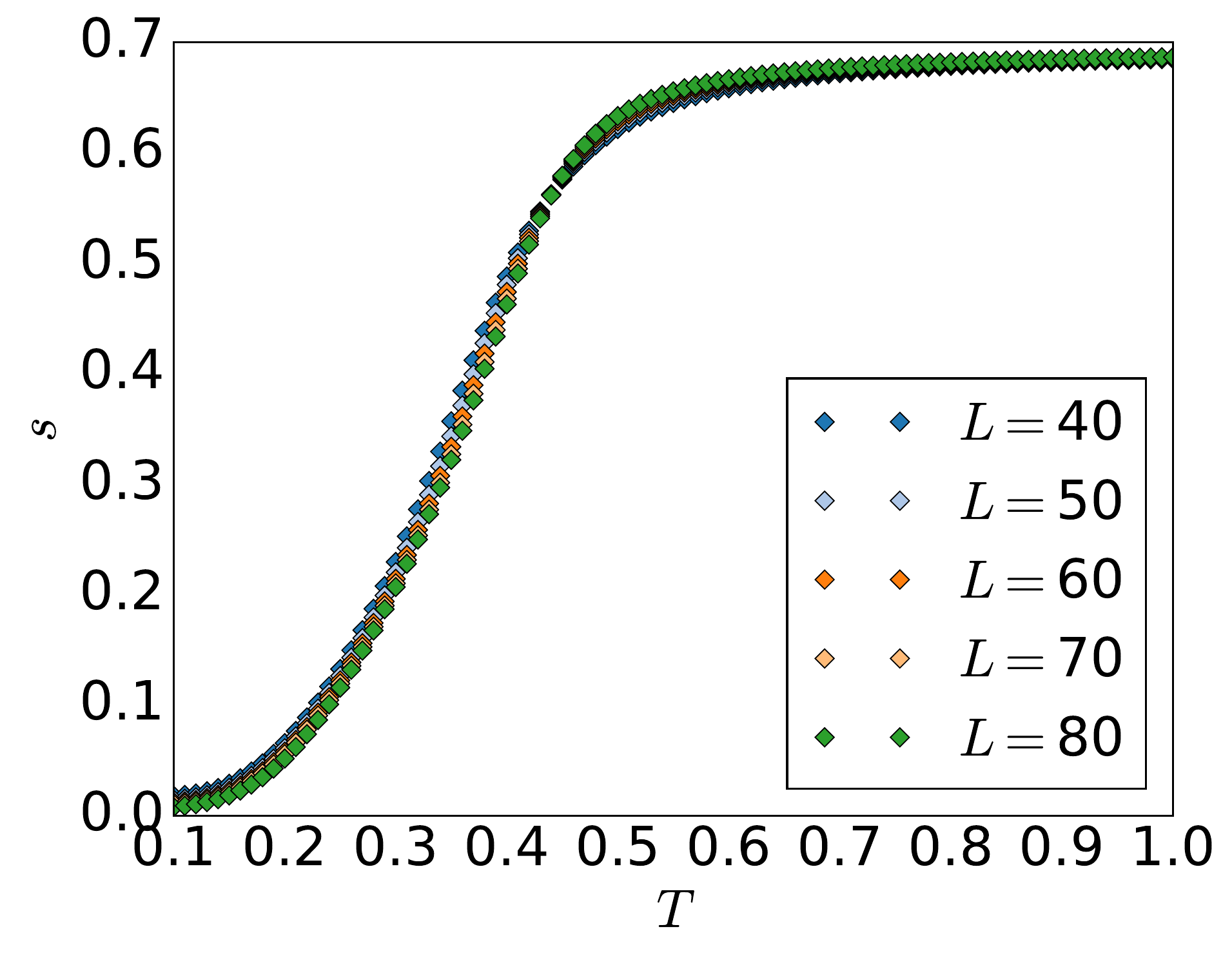}}
	\subfloat[$h=1.2$] {\includegraphics[width=0.25\textwidth]{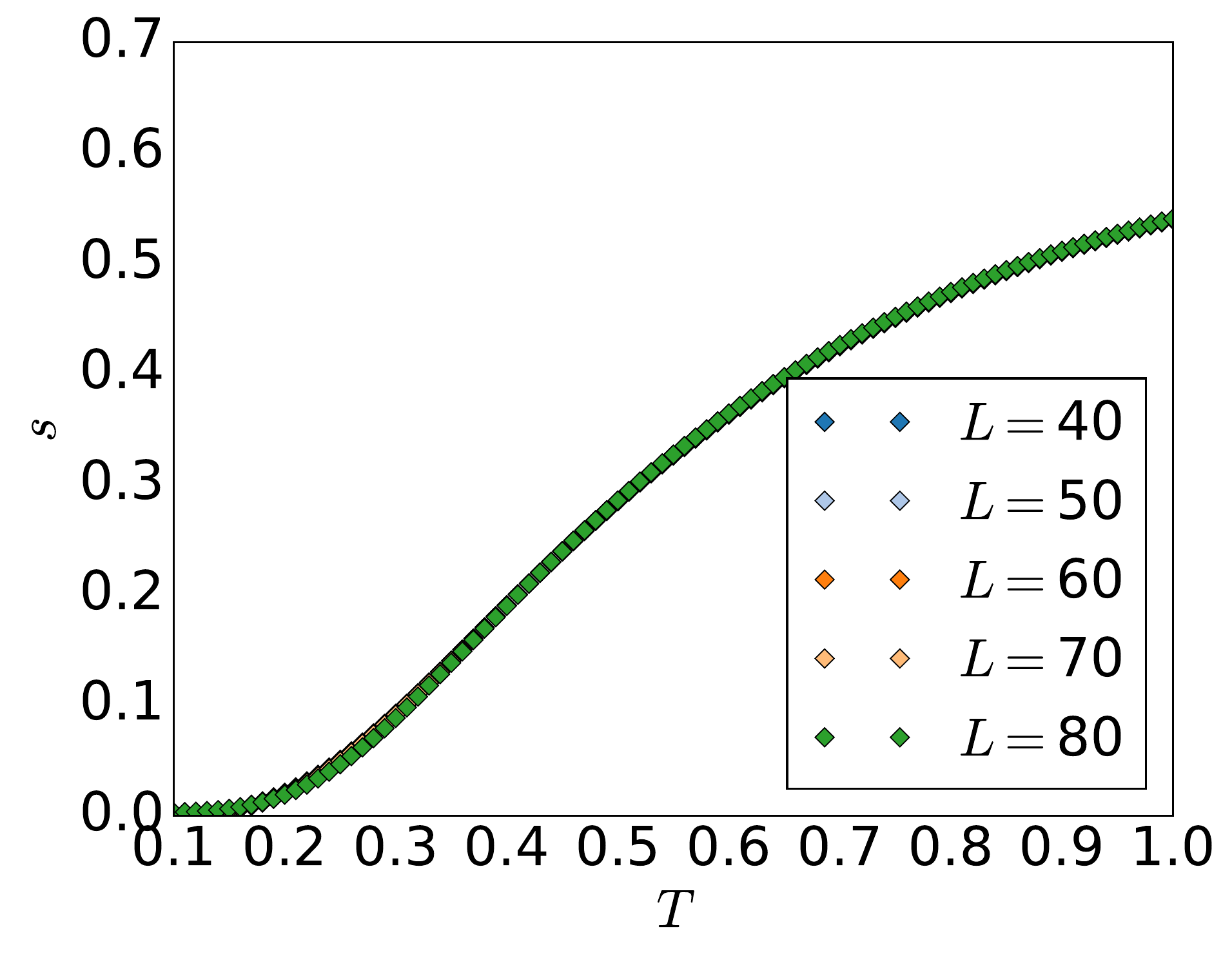}}\\
	\subfloat[$h=0.2$] {\includegraphics[width=0.25\textwidth]{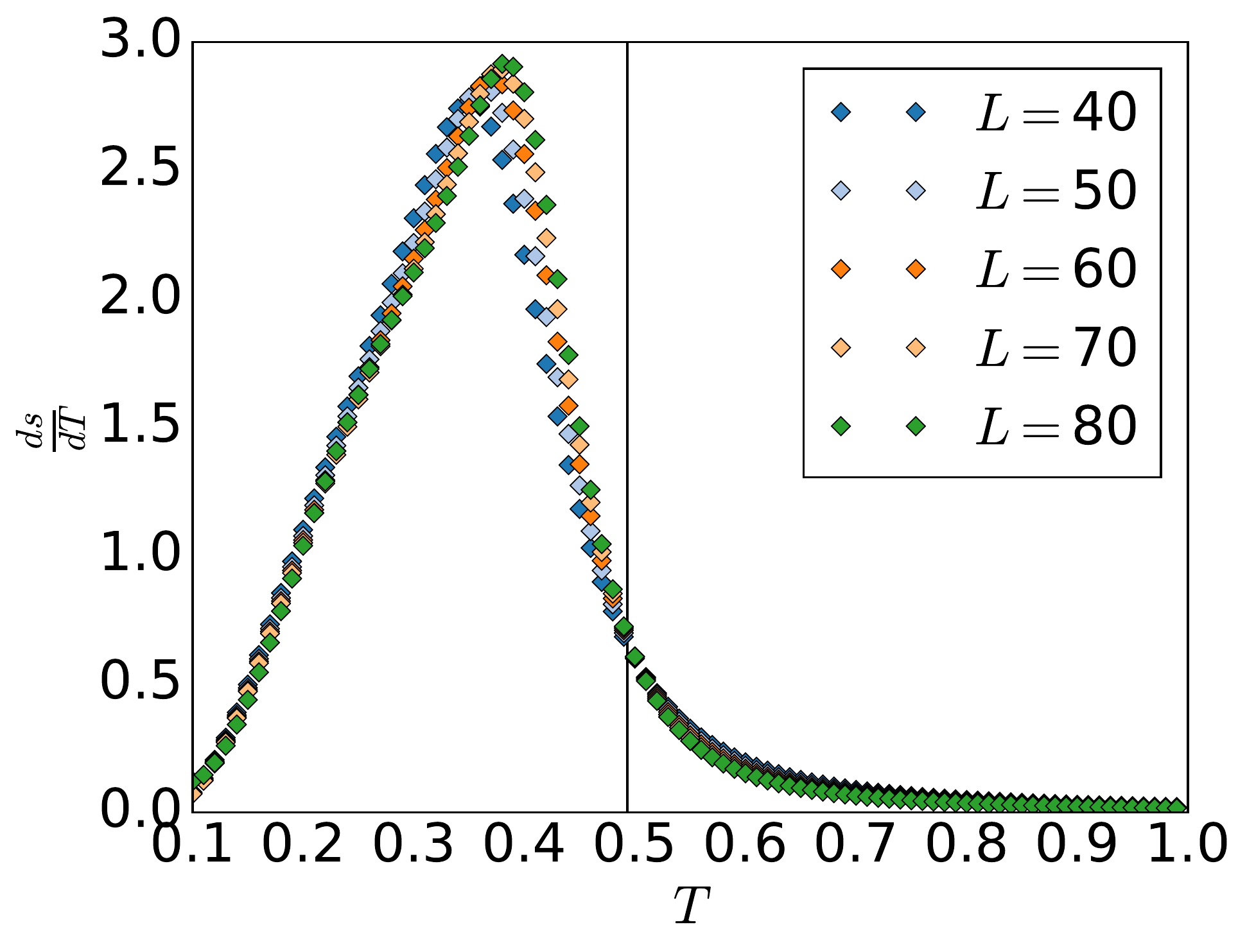}}
	\subfloat[$h=1.2$] {\includegraphics[width=0.25\textwidth]{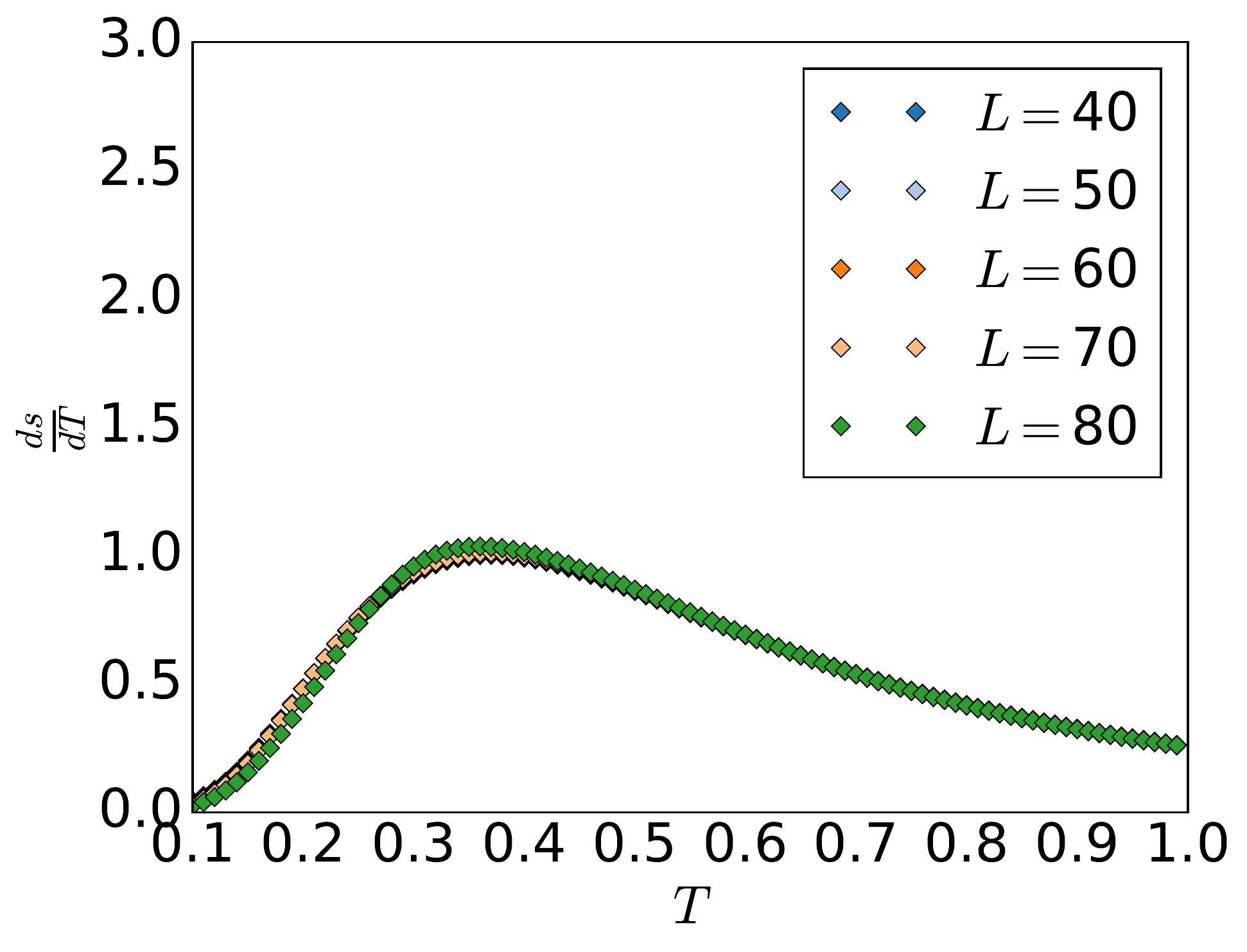}}
	\caption{Top row: von Neumann entropy per site $s$ of the Gibbs state for the LMG Hamiltonian for $h=0.2$ (a) and $h=1.2$ (b) over the temperature $T$ for several system sizes. Bottom row: the discrete temperature derivative of $s$ over $T$ for $h=0.2$ (c) and $h=1.2$ (d).
	}
	\label{entropy}
\end{figure}
\begin{figure}[t]
	\center
	\subfloat[$h=0.2$] {\includegraphics[width=0.25\textwidth]{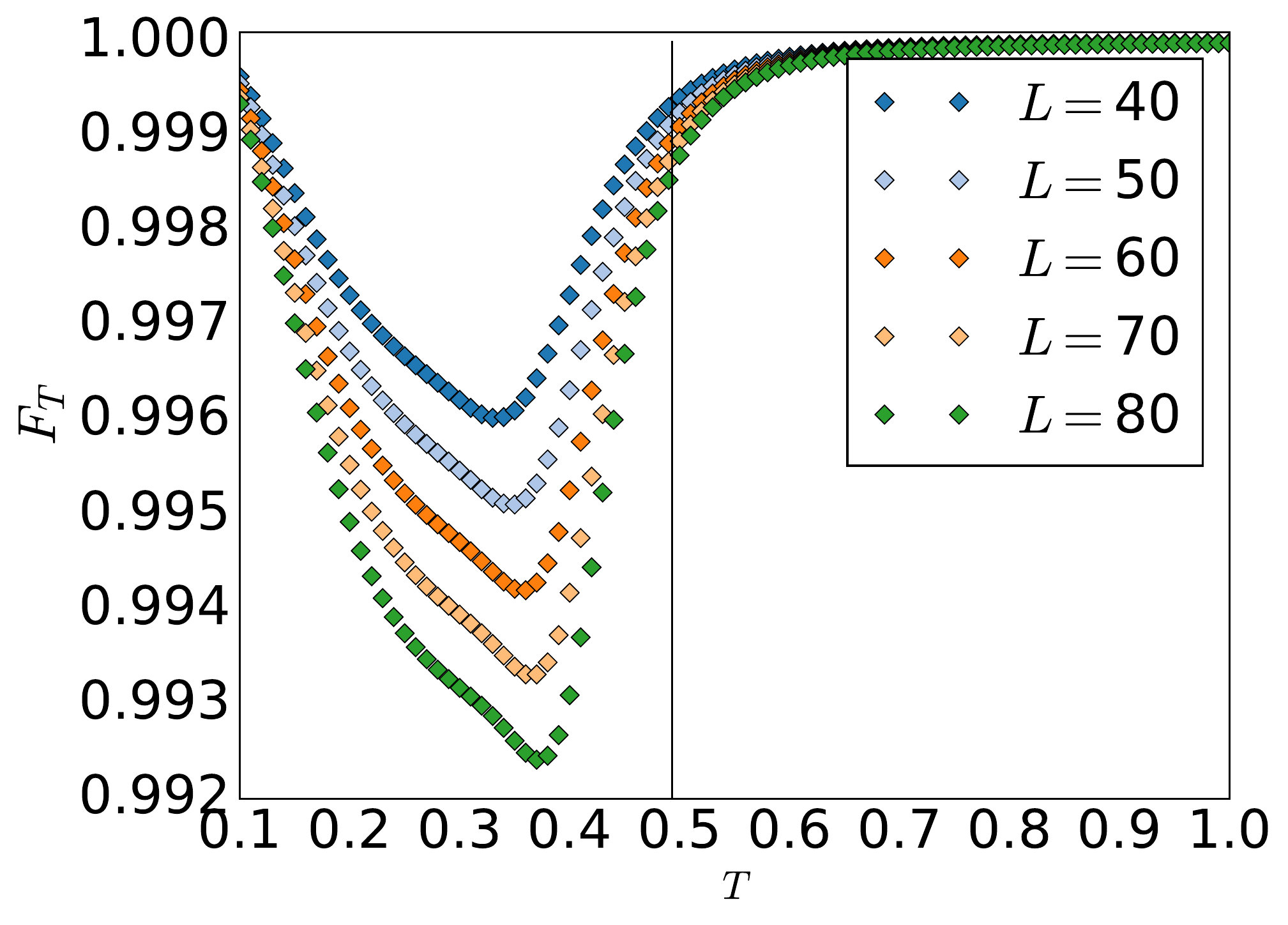}}
	\subfloat[$h=1.2$] {\includegraphics[width=0.25\textwidth]{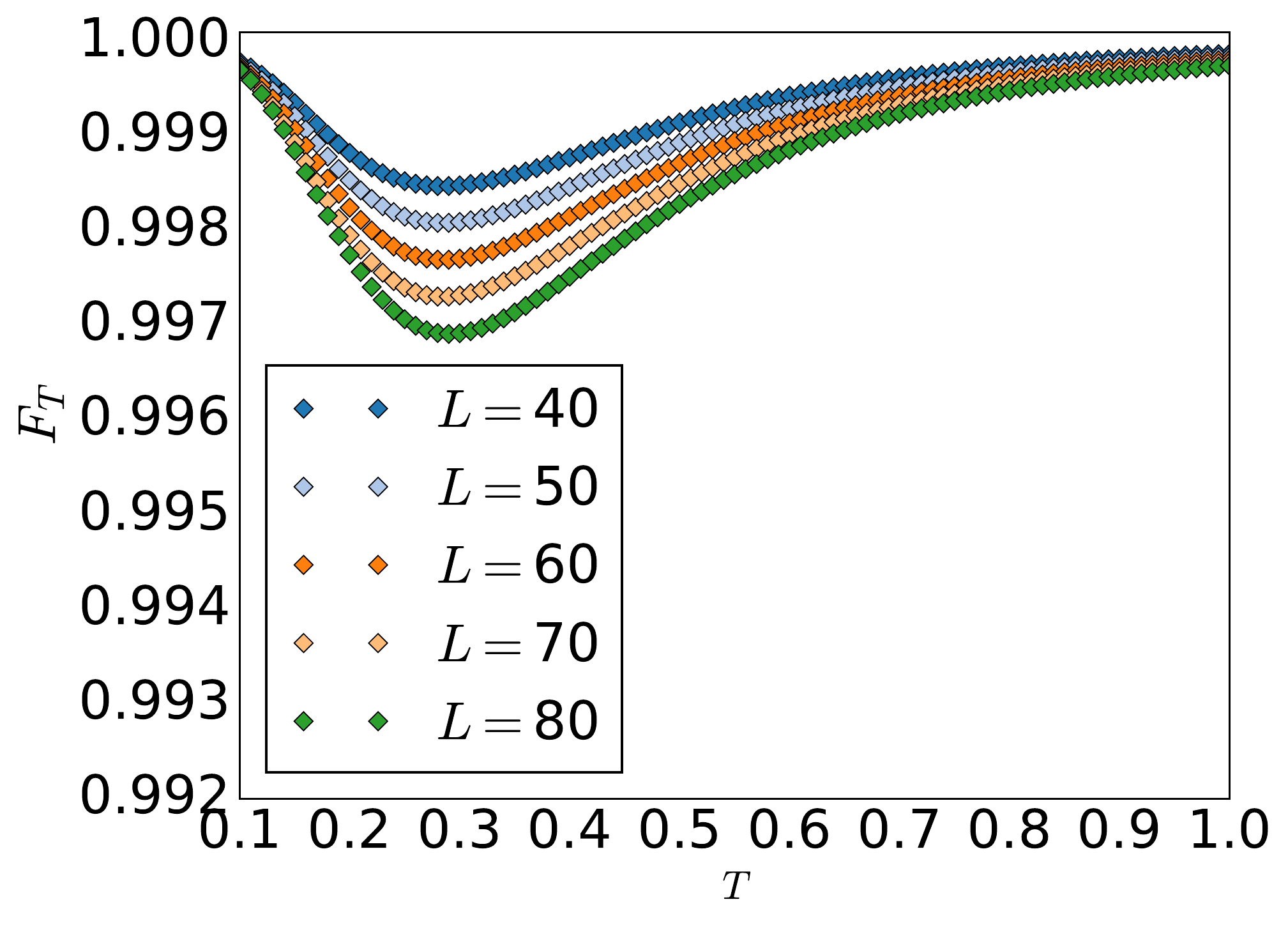}}
	\caption{Thermal fidelity $F_T$ in the Gibbs state for the LMG Hamiltonian, for $h=0.2$ (a) and $h=1.2$ (b) and several system sizes.
	The  black vertical line in Figure (a) indicates the critical temperature $T_c$.
	}
	\label{temp_fidelity}
\end{figure}

\begin{figure}[t]
	\center
	\subfloat[$h=0.2$] {\includegraphics[width=0.25\textwidth]{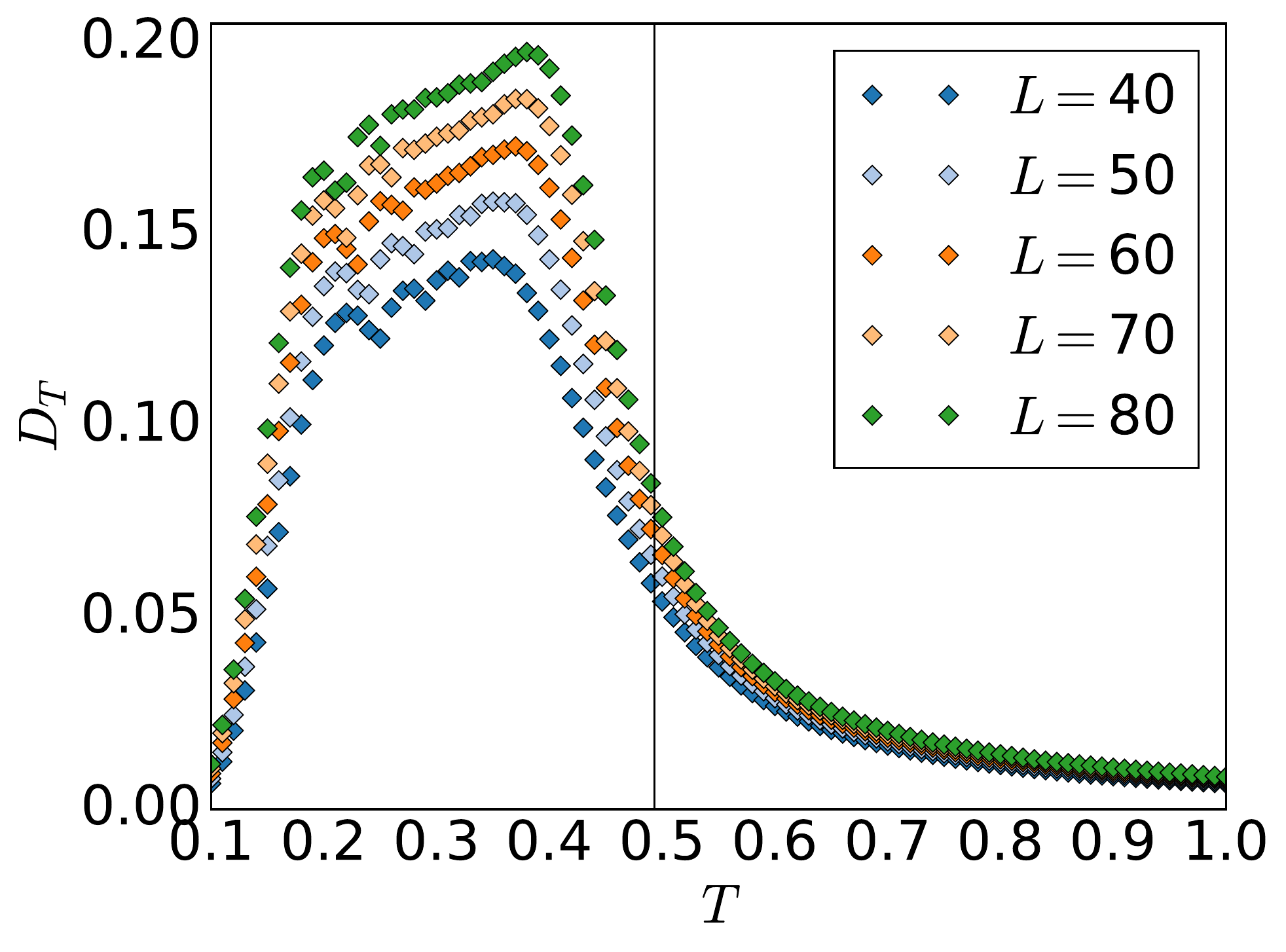}}
	\subfloat[$h=1.2$] {\includegraphics[width=0.25\textwidth]{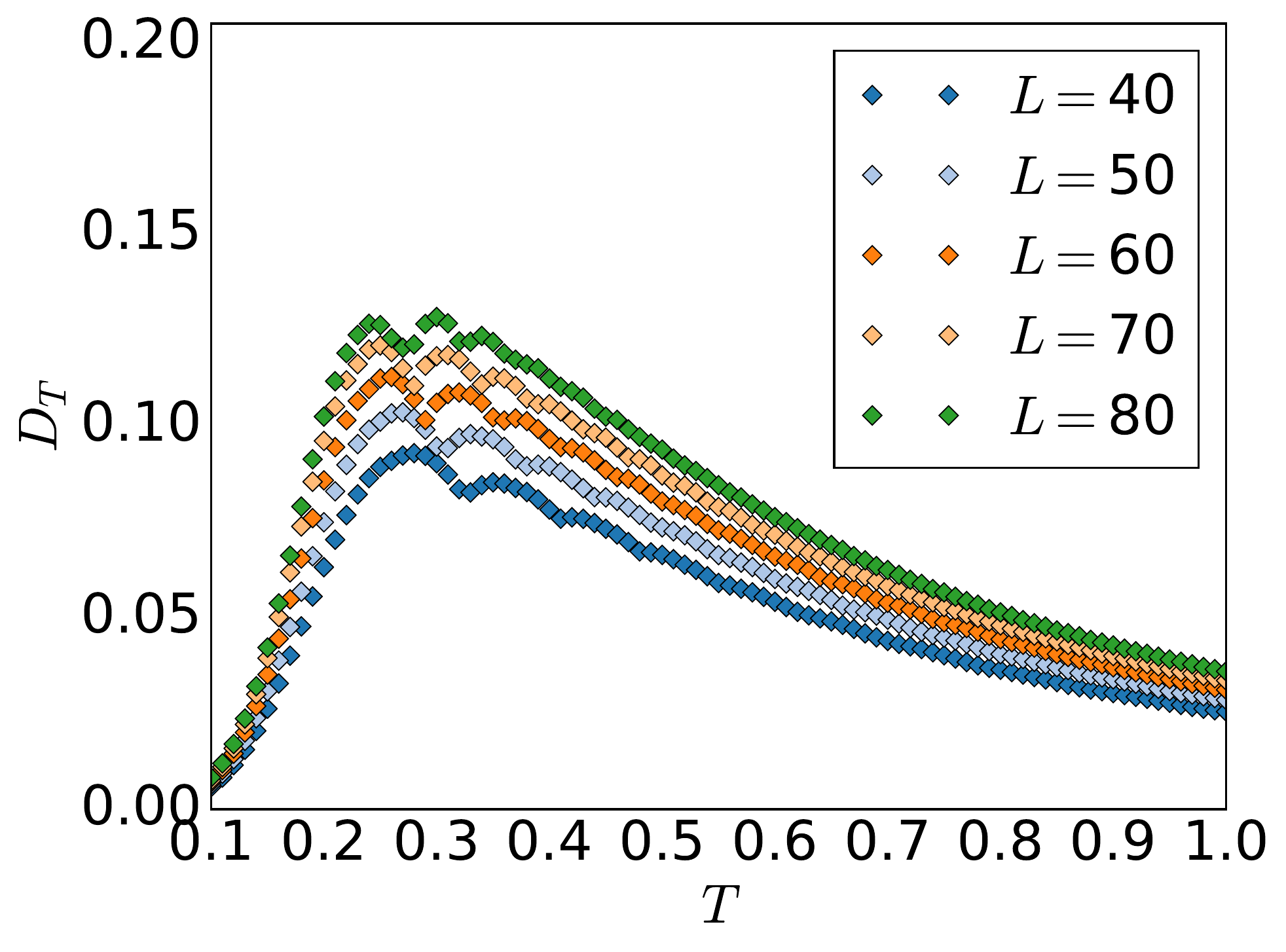}}
	\caption{Trace distance $D_T$ between thermal equilibrium states of the LMG model as a function of the temperature for different system sizes $L$ and $h=0.2$ (a) and $h=1.2$ (b). 
	The black vertical line in Figure (a) indicates the critical temperature $T_c$ in Eq.~\eqref{eq:Tc_LMG}.}
	\label{distance}
\end{figure}


Besides the thermal fidelity, other quantities can be analyzed to locate the critical temperature from 
numerical studies of finite systems. 
Here we use the block Lanczos algorithm described in the previous sections
to compute the specific heat capacity and the entropy density, as well as the thermal fidelity and trace 
distance, and we show how the results are sensitive to the presence of a phase transition.

The Hamiltonian~\eqref{eq:LMG} can be exactly expressed as an MPO with 
constant bond dimension $D=3$.
In contrast, approximating the Gibbs state as an MPO would be difficult, due to the long-range interactions.
Although specific algorithms exist that can approximate the required imaginary time evolution for 
long-range Hamiltonians~\cite{haegeman2011time, zaletel2015time, chen2017se},
the bond dimension required for an accurate description can get large, 
up to the point of not being computable in practice.
Thus, expressing the thermal observables directly as functions of the Hamiltonian
and applying the block Lanczos method to the latter is a more reasonable strategy
for this kind of models.

We consider here finite systems of sizes 
$L \in \{40,\,50,\,60,\,70,\,80\}$ 
and use our method to probe the behavior of 
the quantities of interest
as a function of temperature in the interval  $T \in [0.1,1]$,
 for $h \in \{0.2, 1.2\}$, 
(only for $h<1$ the system exhibits a phase transition).
The maximal size of the Krylov basis is set to 70,
and the maximal bond dimension of the basis MPOs varies from $D=150$ for $L=40$ to $D=250$ for $L=80$.
For the sake of brevity, we will in the following refer to the thermal fidelity at a given temperature $T$,
for states that differ in $\delta T$,
 as $F_T(T):=F_T(\frac{1}{T},\frac{1}{T+\delta T})$ and define the trace distance $D_T(T)$ analogously.
 
Note that for a given system size $L$ and field strength $h$, the algorithm can approximate all 
the required
functions for all values of $T$ in a single run.


Our results are shown in Figures~\ref{heat_cap} (heat capacity per site $c$),~\ref{entropy} (entropy per site $s$),~\ref{temp_fidelity} (thermal fidelity $F_T$) and~\ref{distance} (trace distance $D_T$).
Comparing the plots for both values of $h$ studied,
specially for heat capacity, thermal fidelity and trace  distance, we see a clear signal of the presence of the phase transition for $h=0.2$,
despite
the system sizes considered here being far from the thermodynamic limit.
In particular, the location of the maximum (minimum in the case of $F_T$) can be used to estimate the critical temperature.
Finite size effects are noticeable, with the location of the extremes of all quantities moving closer to the exact 
$T_c$~\eqref{eq:Tc_LMG} for larger system sizes.
We can estimate the location of the critical temperature by a finite size extrapolation, as shown in Figure~\ref{heat_cap}(c).
Using the data of the heat capacity for $L=70,\,80$,  
provides already a relatively good estimation
of the critical temperature  $T_c \approx 0.47 \pm 0.04$,
where the error corresponds to the difference with respect to the estimator obtained from including also $L=60$ in the fit.
The extrapolated value is already close to the exact solution, and compatible with it within the error bars.
To obtain a more precise estimate of the critical temperature, more data for larger systems sizes and possibly a smaller temperature step could be computed with our method.


The von Neumann entropy per site does not show a sharp peak, but it also exhibits a qualitatively different behavior for both 
values of $h$. For $h=0.2$, where a phase transition exists, the value of the entropy per site of the thermal equilibrium state 
develops a rapid change from $s=0$ at low temperatures, to $s\approx 0.7$ after the critical temperature, with the  
increase becoming faster with growing system sizes.
In contrast, in the case $h=1.2$ the entropy density increases smoothly and does not show a clear finite size scaling.
This different behavior is made more apparent by looking at the (discrete) derivative of the entropy with respect to the temperature,
shown in Figures~\ref{entropy}(c) and \ref{entropy}(d). In these plots, similar to the behavior of the heat capacity, we appreciate a sharp peak 
signaling the phase transition only in the case $h=0.2$.


Overall the results show that already with moderate computational effort, our method can be used to 
reveal interesting physical phenomena that would otherwise be hard to access numerically.
As the method is theoretically guaranteed to converge to the exact solution in absence of numerical and approximation errors, more accurate results can in principle be obtained by increasing the maximal bond dimension and size of the Krylov basis.
Indeed, due to the permutation symmetry of the Hamiltonian~\eqref{eq:LMG},
in this particular model large systems can be explored with exact diagonalization, as was 
done in~\cite{quan2009quantum,wilms2012finite}.
The goal of the calculations presented above was thus not to compete with those results, 
but to use them as reference and to probe the performance of the algorithm.
We expect that the strategy presented here will be most useful for 
cases where the dimension of the problem is genuinely exponential,
and exact diagonalization cannot be applied.

\subsection{Two-site correlations}
\begin{figure}[]
	\center
	\subfloat[Absolute values] {\includegraphics[width=0.25\textwidth]{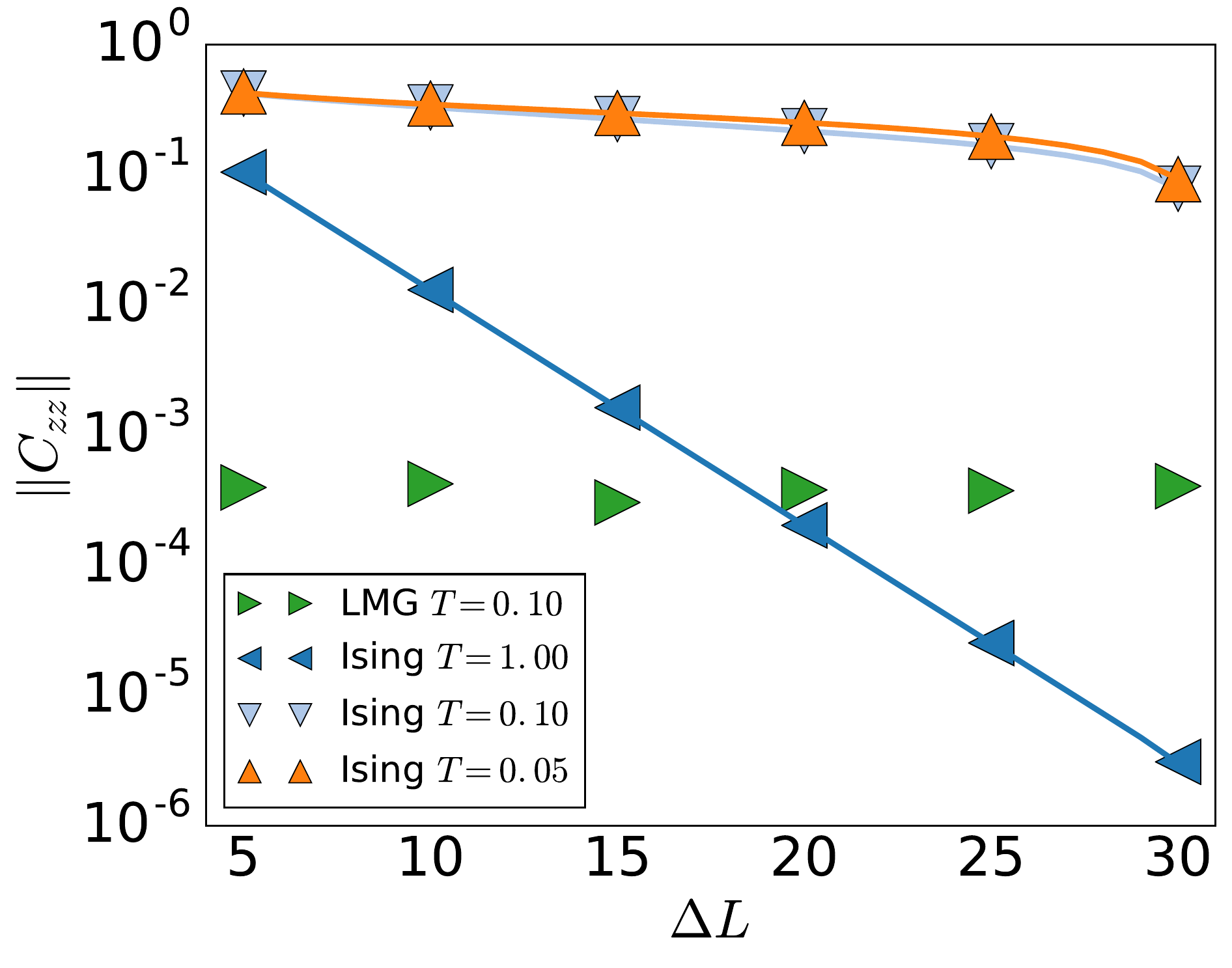}}
	\subfloat[Relative difference] {\includegraphics[width=0.25\textwidth]{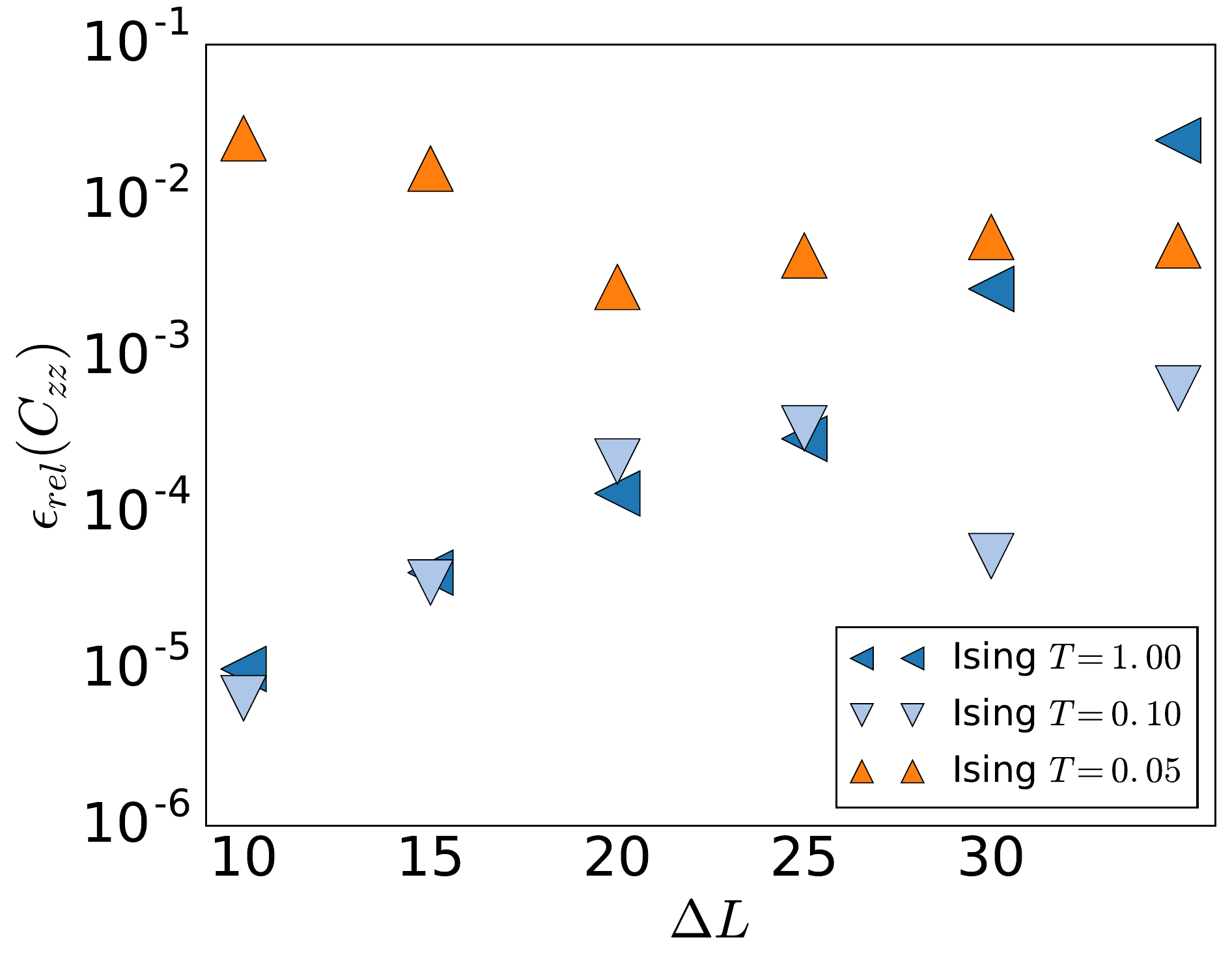}}
	\caption{Two-site correlation function $C_{zz}(L/2,L/2+\Delta L)$ 
	in the thermal state of Ising and LMG Hamiltonians, as a function of the distance $\Delta L$, for 
	a system size $L=60$. 
The discrete data points shown as markers correspond to the results computed with our algorithm, while for the Ising case, quasi-exact results from the MPO approximation are shown as lines, for reference.	
In (a), the absolute values are shown for both Hamiltonians while (b) depicts the relative error for the Ising model.}
	\label{correlations}
\end{figure}

An interesting property of thermal equilibrium states may be the two-site correlations at a certain distance.
If the Gibbs state admits a good MPO approximation
which can be efficiently found,  
for instance via standard imaginary time evolution algorithms,
then such correlations can be accurately computed.
Our method will be most useful when no such MPO representation 
is available, but for the purpose of benchmarking we choose a model where 
a good MPO approximation of the thermal state is easy to find.
In particular, we consider the Ising model in a transverse field on finite open chains,
\begin{align}
H_I = J \sum_{i=1}^{L-1} \sigma_i^{x} \sigma_{i+1}^{x} + g \sum_{i=1}^{L} \sigma_i^{z},
\label{eq:Ising}
\end{align}
and use the standard approximation of the Gibbs ensemble as an MPO 
as a quasi-exact reference to benchmark the estimations obtained with our method.
We will also illustrate the results obtained for the thermal states of the 
LMG Hamiltonian discussed above, 
although in that case no MPO approximation is available for comparison.



We define a two-point correlation function $C_{zz}(i,j) = \braket{\sigma^z_i\sigma^z_j}-\braket{\sigma^z_i}\braket{\sigma^z_j}$.
In the case of a mixed state $\rho$,
\begin{align}
	       C_{zz}(i,j) &= \trace \left(\rho\, \sigma^z_i\sigma^z_j\right)-\trace \left(\rho\, \sigma^z_i\right)\trace \left(\rho\, \sigma^z_j\right),
\end{align}
which yields three expectation values that need to be computed or approximated.
We have seen in Section~\ref{sec:approximation_of_gibbs_states} that our method can 
approximate expectation values 
of positive operators $O$, if there is an efficient MPO representation for $\sqrt{O}$.
Although $\sigma_i^z$ and $\sigma_i^z \sigma_j^z$ are not positive,
they can be decomposed using 
$\sigma_i^z=\Pi_{0}^{i}-\Pi_{1}^{i}$ and
\begin{align}
	\sigma_i^z  \sigma_j^z = \left(\Pi_{0}^{i}\Pi_{0}^{j}+\Pi_{1}^{i}\Pi_{1}^{j}\right)-\left(\Pi_{1}^{i}\Pi_{0}^{j}+\Pi_{0}^{i}\Pi_{1}^{j}\right),
	\label{eq:ZZij}
\end{align}
where $\Pi_{s}^i:=\ket{s}\bra{s}$ for $s=0,\,1$ projects the $i$-th site of the chain onto state $\ket{s}$ 
and acts as the identity on the remaining sites. For simplicity we do not explicitly show the identity operators.
Each of the terms within brackets
is a projector, and therefore positive, and has a square root 
(itself)
that admits an MPO-representation with bond dimension $D=1$ 
 for $\sigma_i^z$ or $D=2$ for $\sigma_i^z \sigma_j^z$.
This formulation thus 
provides us with a means of approximating two-site correlations with the block Lanczos method.

In the particular case of the Ising model, the spin-flip symmetry of the Hamiltonian~\eqref{eq:Ising} allows a
further simplification, since we can write
\begin{align}
	\trace \left [ \rho(H_I,\beta) \sigma_i^z \sigma_j^z\right )]=& 2 \left \{\trace \left (\rho(H_I,\beta) \Pi_{0}^{i}\Pi_{0}^{j}\right )\right. \\
	&\left.- \trace \left (\rho(H_I,\beta) \Pi_{1}^{i}\Pi_{0}^{j}\right )\right \},
\end{align}
halving the number of expectation values that need to be approximated and consequentially improving accuracy and runtime.

In Figure~\ref{correlations}, we show 
the results for the correlation
$C_{zz}(L/2,L/2+\Delta L)$, between the middle site and the right half of the chain,
in the Gibbs state  at temperatures $T \in \{0.05,\, 0.1,\, 1\}$,
for a system size $L=60$, $J=1$ and transverse field $g=1$.
The bond dimension was set to $D=250$ and the maximal Krylov dimension to 100.
We compare the results to those from the MPO
obtained by imaginary time evolution with the purification ansatz,
which were obtained using bond dimension $D=120$ for the purification and Trotter step $\delta=0.01$
and checked to be sufficiently converged for this comparison.

Figure~\ref{correlations}(a) 
 shows that the results obtained for the Ising Hamiltonian by the two different methods are in good agreement 
for all the temperature values considered.
The correlations in the thermal equilibrium state decay exponentially with distance, with the 
correlation length becoming larger for lower temperatures, as the quantum critical point at $T=0$ is approached.
The relative errors, shown in Figure~\ref{correlations}(b), in general increase with the distance, which 
we attribute to the fact that the absolute values are indeed smaller, since we observe an approximately constant
absolute error for all distances.
Notice that the results for the LMG case, shown also in Figure~\ref{correlations}(a),
show that the method can capture the correlation of the model despite its long-range interactions.
We observe that for the smallest temperature, $T=0.05$, the error is considerably larger.
We attribute this comparatively larger deviation mainly to an error in the approximation of the partition function as it enters 
all approximated expectation values.
The error reflects the fact that a larger bond dimension and possibly a larger Krylov dimension is required for $T=0.05$ to obtain more accurate estimates.


\section{Conclusion}
\label{sec:conclusion}

In \cite{august2017approximation} we introduced
a block Lanczos method for approximating functions of the form $\trace \left[f(A)\right]$ 
of any Hermitian operator $A$ given as an MPO.
The method gives access to global functions of the operator that depend on the full spectrum and are usually not accessible with
standard tensor network tools.
In this work we have discussed how to use the method for physically interesting quantities, such as the von Neumann entropy or the trace norm of an MPO, and in particular
how to most efficiently approximate thermal properties.
As we have argued, by 
expressing the quantities of interest in the thermal equilibrium state as functions of the Hamiltonian,
and applying the Lanczos method directly to the latter,
we can explore the thermal properties in an efficient way.
The strategy allows the evaluation of a variety of properties without 
the need of a prior approximation of the thermal state as an MPO,
and can be applied for long-range interactions.
We have discussed how a single run of the algorithm is enough to evaluate different functions 
over the whole range of possible temperatures.


We have then shown how to approximate several physical quantities for Gibbs states, 
namely the heat capacity, thermal fidelity, trace distance, von Neumann entropy and some expectation values.
To illustrate the performance of the method we have presented results for the thermal states of the 
Lipkin-Meshkov-Glick and Ising Hamiltonians.
In the LMG model, we have shown how the method can estimate multiple quantities that detect the 
presence of a thermal phase transition.
The case of the Ising model has been used to benchmark the approximation of 
correlation functions.

This algorithm provides a new tool to extend the capabilities of the tensor network toolbox,
by giving access to global functions whose calculation would  otherwise be unfeasible. 
The calculations presented in this paper aim at benchmarking the method with known results,
but we expect the technique to be most useful in cases when an MPO approximation of the 
Gibbs state is not available, but the Hamiltonian has an MPO description, as can be the case for long-range interactions. 

Apart from algorithmic improvements such as leveraging symmetries in the input to reduce the required bond dimension and runtime, an interesting and promising direction of future research would be to identify additional physical applications of the algorithm that so far were out-of-reach for tensor network algorithms. 
The algorithm's generality and capability of approximating global quantities make this seem like a worthwhile endeavor.


\acknowledgments{
This work was partly funded by the Elite Network of Bavaria via the doctoral programme \emph{Exploring Quantum Matter (ExQM)}.

}

\bibliography{ref}

\end{document}